\documentclass[preprint,showpacs,aps]{revtex4}

\begin{document}

\title{Large voltage from spin pumping in magnetic tunnel junctions}
\author{S. T. Chui, Z. F. Lin}
\affiliation{Bartol Research Institute and Department of Physics and Astronomy,
University of Delaware, Newark, DE 19716}
\date{\today}

\begin{abstract}
We studied the response of a ferromagnet-insulator-normal metal 
tunnel structure under an external oscillating radio frequency
(R.F.) magnetic field. The
D. C. voltage across the junction is calculated and is found not to decrease
despite the high resistance of the junction; instead, it is 
of the order of $\mu V$ to $100\mu V$, much larger than the
experimentally observed value (100 nano-V)
in the "strong coupled" ohmic 
ferromagnet-normal metal bilayers. This is consistent
with recent experimental results in tunnel structures, where the voltage
is larger than $\mu V$s.
The damping and loss of an external RF  field in this structure is calculated.
\end{abstract}

\maketitle





       There has been much recent interest in the spin dynamics in hybrid
nanostructures composed of ferromagnetic and normal metal layers.[1-5]
Following earlier work on the spin torque effect, the spin pumping
effect\cite{sp_theo} has been demonstrated for ''strongly coupled''
ohmic metallic multilayers as an additional contribution to the FMR
linewidth in FM/NM multilayers (where NM is Pt, Pd,
Cu, etc.)\cite{sp_damp_expt}  and more recently
as a difference in voltages of the order of 100 nV between two FM/NM interfaces
of a NM1/FM/NM2 trilayer.\cite{sp_v_expt}
       Two types of metallic structures are commonly studied.
In addition to the "strongly coupled" ohmic multilayer systems,
"weakly coupled" tunnel structures have also been extensively studied.
The physics of these two types of systems can be very different.\cite{chui1}
In particular, for the tunnel structures, the coupling of the longitudinal
magnetization and the charge lead to magnetization and charge dipole
layers at the interface. After the effect of the electron-electron
interaction is included, it is found that because of the large difference
of the length scales associated with the charge (screening length, $\approx
1\AA$) and
the spin (spin diffusion length, $\approx 100\AA$) fluctuations,
there is a larger splitting of the chemical potentials than that predicted 
by the conventional spin accumulation picture.\cite{past} Whereas the 
conventional picture suggests that the splitting scales with the current 
and will decrease with an increase with resistance, this is no longer
true in the more complete picture.

       Recently Moriyama and coworkers\cite{expt} reported
measurements of the dc voltage attributed to the spin pumping effect in
different tunnel junctions, and demonstrate that the voltage is
larger than micro-volts, enhanced orders of magnitude compared
to that for metallic trilayers.   In this paper, we generalize our recent
work on spin torque\cite{st} to the spin pumping situation and found
an enhanced voltage for the tunnel structures, in agreement with the
experimental results. We now describe our results in detail.

        The system we have in mind is a ferromagnet-normal metal
tunnel junction where the
two interfaces between the ferromagnet-insulator-metal sandwich
structure are assumed to be at $z=\pm d/2$. We assume the $z$ axis to be
perpendicular to the faces of the tunnel junction. The initial magnetization
is assumed to be in the $x$-$y$ plane with an orientation given by $\mathbf{%
p}_{0}^{L}= \mathbf{e}_{x}$ for the ferromagnet on the left hand
side of the sandwich structure.

Because the work functions of the metals on opposite sides
of the junction may not be equal, at zero external radio frequency (RF) 
field there will be a
charge dipole layer formed at the interfaces. What we are calculating here
are the changes from the zero field situation.
This surface inhomogeneity can lead to an additional contribution to
the increase
in the FMR damping, as we explain below. The experimental structures
usually possess edge domains where the switching starts. The magnetization
is thus not completely uniform in the $x$-$y$ plane. To bring out the
essential physics, we shall not consider this complication in the present
paper but we hope to come back to this in the future.

Under an external time varying RF field, we expect
the magnetization in the ferromagnet to be a sum
of a uniform magnetization which is a solution of the inhomogeneous
Bloch (Landau-Gilbert) equation due to the external field and a spatially varying
solution of the homogeneous equation so that the boundary condition can be
satisfied. This spatially varying part provides for the additional damping 
and the voltage observed in the experiments. Our approach is to obtain general
solutions in each part of the junction (Eq. (11), (12), (15), (16)). 
The amplitudes of these solutions
are determined by the boundary conditions (Eq. (7)). From these amplitudes, the
voltage and the damping can be derived. We first describe the general
solution of the magnetization in a ferromagnet.

{\it Magnetization in a ferromagnet:}
Our starting point is the equation of motion of the charge and the
magnetization. For the charge, it is just the equation of charge current
conservation
\begin{equation}
\nabla\cdot\mathbf{J}_e=-\frac{\partial \delta\! n}{\partial t}
\end{equation}
where $\mathbf{J}_e$ is the total current. The equation for the
magnetization $\mathbf{M}$ has been much discussion extensively in the past.%
\cite{EB} The equation takes the form of the phenomenlogical classical
Landau-Lifshitz (Bloch) equation with longitudinal and transverse damping and an additional source term
\begin{equation}
\frac{\partial \mathbf{M} }{\partial t} - \gamma \mathbf{M}\times\mathbf{H}
-\alpha \mathbf{M}\times(\mathbf{M} \times\mathbf{H})+ \nabla\cdot{\hat{%
\mathbf{J}}_{\mathrm{M}}} = -\frac{\delta\!\mathbf{M}}{\tau}
\end{equation}
where $\gamma$ is the gyromagnetic ratio, and $\mathbf{H}$ is the effective
field describing the precession of the magnetic moments given by $\mathbf{H}=%
\mathbf{H}_{\mathrm{e}}+\mathbf{H}_{\mathrm{an}}+\mathbf{H}_{\mathrm{dip}}+%
\mathbf{H}_{\mathrm{ex}}.$ $\mathbf{H}_{\mathrm{ex}}=J\nabla^2\mathbf{M}$ is
the effective field due to direct exchange;
the anisotropy term includes a
bulk and a surface
anisotropy energy
$\mathbf{H}_{\mathrm{an}}=
H_{ab}+H_{as},$ $H_{ab}
=K \mathbf{M}_0,$
$H_{as}=K_sM_s$ where $M_s={\bf M}_0\delta(z+d/2).$
Here we have separated a bulk and a surface contribution that acts
on the surface magnetization $M_s$. For simplicity
we have assumed this surface contribution localized at the interface.
The other terms can also contain a surface contribution and can be
treated in a similar manner as this ansiotropy contribution. For simplicity
of presentation we illustrate our results with just this term.
$\mathbf{H}_{%
\mathrm{e}}$ represents the external field; and $\mathbf{H}_{\mathrm{dip}}$
denotes the dipole-dipole interaction. 
${\hat{\mathbf{J}}_{\mathrm{M}}}$ is a spin current (tensor). The
currents are driven by density gradients (diffusion) and external forces.
\begin{equation}
\begin{array}{l}
\mathbf{J}_e=-\sigma\nabla V -e D\nabla \delta\! n -\displaystyle D_M \nabla
(\Delta\!\mathbf{M}\cdot \mathbf{p}_0) \\
{\hat{\mathbf{J}}_{\mathrm{M}}}=- \sigma_M \nabla ( V \mathbf{p}_0) -
D_M^{\prime }{\nabla\Delta\!\mathbf{M}} - D^{\prime }\nabla ( \delta\! n
\mathbf{p}_0)%
\end{array}
\label{JeJM}
\end{equation}
where 
$\sigma$, $\sigma_M$ are the effective conductivities for the charge and
magnetization. 
$\mathbf{p}_0$ is a unit vector along the direction of the equilibrium
magnetization: $\mathbf{p}_0=\displaystyle\frac{\mathbf{M}_0}{|\mathbf{M}_0|}
$ with $\mathbf{M}_0$ the local equilibrium magnetization, $\Delta\!\mathbf{M%
}=\mathbf{M}(1-|\mathbf{M}|/M_0)$ is a change in magnetization. $D$, $%
D^{\prime }$, $D_M$, $D_M^{\prime }$ are the effective diffusion constants. $%
V=V_e+W$, with $V_e$ the electric potential describing the external electric
field and $W$ the local electric (screening) potential due to the other
electric charges determined self-consistently by
\begin{equation}
W(\mathbf{r})=\int d^3\mathbf{r}^{\prime }U(\mathbf{r}-\mathbf{r}^{\prime })
\delta\! n(\mathbf{r}^{\prime })
\end{equation}
with $U$ the Coulomb potential. The total number density of charge carriers
and $x$ component of magnetization are given by $n=\sum_s n_s,
M_{x}=\sum_s s n_s$. In the coordinate system with one of the
coordinate axis along the direction of the magnetization, the spin
current can be understood as the difference of the spin up current and the spin
down current. The vector dependence is such that the equation is covariant.
The Landau-Liftshitz equation
without the source term $\nabla \cdot J_M$ is believed to describe the
physics of ordinary domain walls where the direction of the magnetization
changes but its magnitude remains fixed. Eq. (3) is consistent with this
belief. For ordinary domain walls, $J_M=0$. $\tau$ is the longitudinal relaxation time,
describing the relaxation of the system towards its local equilibrium value
of magnetization. $\alpha$ measures the transverse (Gilbert) damping term.

Substituting the expression for ${\hat{\mathbf{J}}_{\mathrm{M}}}$ into the
modified Landau-Lifshitz 
equation (2) we obtain the linearized relaxation equation for $\mathbf{M}$:
\begin{equation}
\nabla^2\delta\!\mathbf{M} - (\displaystyle\frac{1}{l_{sf}^2}+i\omega
/D_M^{\prime }) \delta\!\mathbf{M} +\zeta\mathbf{p}_0\times(\nabla^2\delta\!%
\mathbf{M}-\frac{\kappa}{l_{sf}^2}\delta\!\mathbf{M} -\kappa_s
\delta {\bf M}_s/l_{sf}^2)
+\gamma(\delta \mathbf{%
M}\times \mathbf{H}_0+\mathbf{M}_0\times \mathbf{H}_1)  \label{eq1}
\end{equation}
\[
+\alpha\delta \mathbf{M}_{\perp}M_0H_0/D_M^{\prime }= -(D^{\prime }/D_M^{\prime })
\mathbf{p}_0 ( \nabla^2\delta\! n -\frac{\delta\! n}{\lambda_0^2})
\]
where only $\mathbf{H}_{\mathrm{ex}}=J\nabla^2\mathbf{M}=J\nabla^2\delta\!%
\mathbf{M}$ and $\mathbf{H}_{\mathrm{an}}=K\mathbf{M}_0$ are kept in the
precession term $\gamma \mathbf{M}\times \mathbf{H}$, and use has been made
of Gauss' law: $\nabla^2V=\nabla^2W=-\displaystyle\frac{e}{\epsilon_0}%
\delta\!n$. The bare spin diffusion length $l_{sf}$ and the bare screening
length $\lambda_0$ are given by $l^2_{sf}=\tau D_M^{\prime }\ \
\mbox{\rm
and} \ \ \lambda_0^2=\frac{\epsilon_0 D^{\prime }}{\sigma} $ respectively.
Other dimensionless parameter are $\zeta=\displaystyle\frac{\gamma|\mathbf{M}%
_0| J}{D_M^{\prime }}$ and $\kappa=\displaystyle\frac{l_{sf}^2 K}{J}$.


The charge current conservation (1) yields, for the steady state without
linearization,
\begin{equation}
\frac{1}{\lambda^2_0}\delta\! n - ( \nabla^2 -i\omega /D) \delta\! n
-(D_M/D) \nabla^2 (\delta\! \mathbf{M} \cdot \mathbf{p}_0)=0 ,  \label{eq2}
\end{equation}
which, together with Eq.(\ref{eq1}), describes the distribution of the
charge and magnetization away from the tunnel junction in
terms of their values at the junction. (To simplify the algebra, we have
made the approximation that D=D'). The values of the charge and
magnetization densities at the junction can be determined by matching
boundary conditions across the barrier. We first solve these equations in
the metal part of the junction. These solutions determine the charge and
magnetization dipole layers.

The solution of eq. (\ref{eq1}) can be written a sum of two terms, 
$$\delta {\bf M}=\delta {\bf M}_0
+\delta {\bf m}^i,$$ 
a  spatially
uniform ($\delta {\bf m}^i$) solution of
the bulk inhomogeneous equation with the source
term $\mathbf{M}_0\times \mathbf{H%
}_1$ and a sptially varying solution ($\delta {\bf M}_0$) of the homogeneous equation.
The inhomogeneous bulk equation is
\[
- (\displaystyle\frac{1}{l_{sf}^2}+(i\omega-\alpha^{\prime }) /D_M^{\prime
})\delta\mathbf{m} -\zeta\mathbf{p}_0\times\frac{\kappa}{l_{sf}^2}\delta\!%
\mathbf{m} +\gamma(\delta \mathbf{m}\times \mathbf{H}_0+\mathbf{M}_0\times
\mathbf{H}_1) =0
\]
where $\alpha^{\prime }=\alpha M_0H_0.$ This is the conventional FMR
equation, which can be readily solved. Define ${\bf e}_{\pm}={\bf
e}_z\pm i{\bf e}_y,$ then ${\bf e}_x\times {\bf e}_{\pm}=\pm i {\bf
e}_{\pm}.$ We write the transverse magnetization as
$\delta {\bf m}_{\perp}^i=\sum \delta m_{\pm}{\bf
e}_{\pm}$ and obtain $$\delta m_{\pm}=\chi^0_{\pm}H_{1,\pm}$$ where
$1/\chi^0_{\pm}= [(-\pm
i\displaystyle\frac{1}{l_{sf}^2}+(i\omega-\alpha^{\prime })
/D_M^{\prime })+\zeta\frac{\kappa}{l_{sf}^2}+\gamma H_0]/\gamma
M_0.$
Associated with this transverse magnetization,
there is a change of the {\bf longitudinal} magnetization
given by $$\delta m_x^i=M_0-(M_0^2-\delta
m_{\perp}^{i2})^{1/2}\approx 0.5\delta m_{\perp}^{i2}/M_0.$$
This is the lowest order correction to the longitudinal magnetization. Higher
order nonlinear corrections to the transverse magnetzation will produce changes in the longitudinal component that is higher than 3rd order in $H_1.$
In the equation of motion (2), no lower order correction are produced.

The equation for the spatially varying term becomes
$$\nabla^2\delta\!\mathbf{M_0} - (\displaystyle\frac{1}{l_{sf}^2}+i\omega
/D_M^{\prime }) \delta\!\mathbf{M_0} +\zeta\mathbf{p}_0\times[\nabla^2\delta\!%
\mathbf{M_0}-\frac{\kappa}{l_{sf}^2}\delta\!\mathbf{M_0}
-\kappa_s(\delta {\bf m}^i+\delta {\bf M}_{0s})/l_{sf}^2]
+\gamma \delta \mathbf{%
M}_0\times \mathbf{H}_0
$$
$$
+\alpha \delta \mathbf{M}_{0\perp}M_0H_0/D_M^{\prime }= -(D^{\prime
}/D_M^{\prime }) \mathbf{p}_0 ( \nabla^2\delta\! n -\frac{\delta\!
n}{\lambda_0^2})
$$
The solution of this equation is similar to that in our previous
studies.\cite{st} Away from the boundary, the surface terms are absent. This
equaion becomes homogeneous.
We solve this homogeneous equation and form linear
combinations of them to satisfy the boundary constraints.
By integrating this equation over a small region of space at the
boundary we arrive at the condition that the difference between the tunnelling 
and the ferromagnet pseudo spin current is equal to surface
anisotropy term:
\begin{equation}
J_{\bf M}^t-I_{\bf M}^L=\gamma K_s{\bf M}_0\times (\delta {\bf m}^i
+\delta {\bf M}_{0s}).
\end{equation}
where $I_{\bf M}^t$ is the tunnelling magnetization current,
the pseudo spin current\cite{BJZ} 
$ J_{\bf M}=J_{\bf M}-\gamma J{\bf
M}_0\times \partial_z\delta {\bf M}$ includes an extra term involving
the exchange that affects only the transverse magnetization current.
We expect this extra term to be also present for ohmic junctions
but so far it has not been included.
In previous spin pumping studies on 
ohmic junctions, a term of a similar functional form $g{\bf n}\times\partial 
{\bf n}/\partial t$ (${\bf n}={\bf M}/|M|$) has been discussed. 
However, the coefficient was interpreted as a spin mixing conductance. 
We next discuss the solution of the homogeneous equation.

We expect the charge and magnetization dipole layers to decay away
from the interface with length scales controlled by the spin diffusion
length and the screening length. Because of the vector nature of the
magnetization, there are three normal modes by which they can decay away
from the interface. Including the charge degree of freedom, there are four
normal modes that one can consider. For the ferromagnetic metal on the left
hand side, we thus consider the following ansatz:
\begin{equation}
\delta\! n^L=\sum_{i=1}^4 \delta\! n_{i0}^L e^{(z+\frac{d}{2})/l_{i}},\ \
\delta\!\mathbf{M}_0^L=\sum_{i=1}^4 \delta\! \mathbf{M}_{i0}^L e^{(z+\frac{d}{%
2})/l_{i}},  \label{ansatz}
\end{equation}
where the superscript $L$ denotes the left hand side.

Letting the coefficients before the exponential scaling functions vanish for
steady-state solutions, we get
for small $\omega$
the renormalized screening length
\begin{equation}
l_1=\lambda_0 \xi_1^{1/2},
\end{equation}
the renormalized spin diffusion length
$$l_2=l_{sf}\xi_2^{1/2},$$
and a combination of the exchange length and the spin diffusion length
$$l_{3,4}
= l_{sf}/[(1-\pm i\zeta\kappa_r)/(1-\pm i\zeta)+
(i\omega-\alpha')l_{sf}^2 /(D_M'^2(1-\pm i\zeta))]^{1/2}.$$
The $\xi$s and $\beta$ 
are measures of the asymmetry of the spin up and spin down 
conductivities of the ferromagnet:
$\xi_1=[1-D^{\prime }D_M/(DD_M^{\prime })] /
[1-\sigma_MD^{\prime }D_M/(\sigma DD_M^{\prime }) +i\omega \lambda^2_0/D].  $
$\xi_2= (1-\beta^2) /(1-i\omega l_{sf}^2/D_M^{\prime }),$
$\beta^2= [1-D^{\prime }D_M \sigma_M/(\sigma DD_M^{\prime })].$
$\kappa_r=\kappa+\gamma H_0l_{sf}^2/\zeta M_0$. As we shall see below, $l_3$
and $l_4$ correspond to length scales with which the ``precession'' dies
away from the interface. The additional term $\gamma \delta M\times H_0$
modifies these two lengths accordingly. The screening length and
the spin diffusion length are renormalized.
From eq. (6) we find that the charge densities can be related
to the magnetization densities by
\begin{equation}
\delta\! n_{10}^L=e(\xi_1^L-1) \delta\!M_{10}^L/\mu_B, \ \ \ \ \delta\! n_{20}^L=%
\displaystyle\frac{e\lambda_0^2D_M^L}{\mu_B l_2^2D^L}\delta\!M_{20}^L , \ \ \ \
\delta\! n_{30}^L=\delta\! n_{40}^L=0.  \label{n2M}
\end{equation}
Because $l_2>>\lambda_0,$ $\delta n_{20}/e<<\delta M_{20}/\mu_B.$
As we see below, in general $\delta M_{20}$ is much less 
than $\delta M_{10}.$ Inserting
the ``eigen-solutions'' into equations (\ref{ansatz}), we finally obtain
analytic expressions for the dipole layers:
\begin{eqnarray}
\delta\! n^L&=& \delta\! n^L_{10}e^{(z+\frac{d}{2})/l_1} +\delta\!
n^L_{20}e^{(z+\frac{d}{2})/l_{2}}  \label{deltan} \\
\delta\!\mathbf{M}^L&=& \mathbf{p}_0^L \delta\! M^L_{10} e^{(z+\frac{d}{2}%
)/l_1} +\mathbf{p}_0^L \delta\! M^L_{20} e^{(z+\frac{d}{2})/l_{2}} +\mathbf{%
e}_+^L \delta\! M^L_{30} e^{(z+\frac{d}{2})/l_3} +\mathbf{e}_-^L \delta\!
M^L_{40} e^{(z+\frac{d}{2})/l_4}.  \label{deltaM}
\end{eqnarray}
The two transverse modes corresponds to the left and right
circularly polarized modes $\mathbf{e}_{\pm}$. $\delta\! M^L_{i0}$, with $%
i=1,2,3,4$, are to be determined later. Terms of the order $%
(\lambda_0/l_{sf})^2$ or higher have been neglected since $%
l_{sf}^2>>\lambda_0^2$. Also, to simplify the algebra
we have assumed that the ferromagnetic thickness 
$d_F$ to be larger than the spin diffusion length so that we do not need
to worry about "reflection" effects from the leads.
As advertised, the charge dipole layer is the sum of
two terms, one decaying with a length scale of the screening length; the
other, the spin diffusion length. The vector magnetization dipole is now a
sum of four terms. The first two ( $\delta\!\mathbf{M}_{10} ^L$, $\delta\!%
\mathbf{M}_{20} ^L$ ) are along the direction of the original magnetization;
the last two are perpendicular to the direction of the original
magnetization and describes the precession of the magnetization around the
original axis. Again, the first two terms correspond to decay lengths of the
order of the spin diffusion length and the screening length, while the
precession term only decays with a length scale that is a combination
of the exchange length and the spin diffusion length.

With equations (\ref{deltan}) and (\ref{deltaM}), the charge and
magnetization currents $\mathbf{J}_e$ and ${\hat{\mathbf{J}}_{\mathrm{M}}}$
can be worked out as
\begin{eqnarray}
\mathbf{J}^L_e &=& \sigma \mathbf{E}_{ext}  \label{Je0} \\
{\hat{\mathbf{J}}_{\mathrm{M}}}^L &=& \sigma_M \mathbf{E}_{ext} \mathbf{p%
}_0^L/e +  \nonumber \\
& & + \frac{(1- \beta^2) D_M^{\prime }}{l_{2}} \mathbf{e}_z \mathbf{p}_0^L
\delta\! M^L_{20} e^{\delta z/l_{2}} + \frac{D_M^{\prime }}{l_3}\mathbf{e}_z%
\mathbf{e}_+ \delta\! M^L_{30} e^{\delta z/l_3} + \frac{D_M^{\prime }}{l_4}%
\mathbf{e}_z\mathbf{e}_- \delta\! M^L_{40} e^{\delta z/l_4}  \label{JM0}
\end{eqnarray}
where $\delta z=z+\frac{d}{2}$, $\mathbf{E}_{ext}=E_{ext}\mathbf{e}_z$ is
the external electric field inside the conductor.
Note that
the magnetization current is not a function of the rapidly varying part
of the charge and magnetization densities $\delta n_{10},$ $\delta M_{10}$.
In principle,
the magnetization current can contain a term of the form
$J_{M1}\exp(z/l_1).$
In the generalized Landau-Gilbert equation (Eq. (2)), terms of different
functional dependence are each equal to zero. The only terms
that are proportional to $\exp(z/l_1)$
comes from $\nabla\cdot J_M$ and is proportional to $J_{M1}/l_1.$
This term and hence its contribution to the magnetization current
is equal to zero. To match the quantities at the boundaries
we next consider the charge and magnetization in a normal metal (N).



{\it Normal metal:}
On the N side, the charge and magnetization are
not coupled. The charge is given by
\[
\delta n^R=\delta n^R_0\exp(-z/\lambda).
\]
The magnetization satisfies the equation
$\partial_t \mathbf{M}=(D_n\partial^2_z
-1/\tau^N_{sf}) \mathbf{M}=0.$
From this we obtain 
\[
\delta \mathbf{M}^R=\delta \mathbf{M}^R_0 \exp[-(z-d/2)/l_R].
\]
The longitudinal magnetization current at the interface
(z=d/2) is given by
 $J_M^R=-D_N\delta M^R_{0x}/l_R.$

The longitudinal magnetization current at the left
interface is given by eq. (16).
Equating $J_M^R$ to $J_M^L,$ we get
\begin{equation}
\delta M^R_{0x}=-(1-\beta^2)D_M^{\prime L}\delta M^L_{20}l_R/[D_Nl_2].
\end{equation}
The magnetization on the right is proportional to $\delta M^L_{20}$ and is
not a function of $\delta M^L_{10}.$ As we shall see below, $\delta M^L_{10}>>
\delta M^L_{20},$ hence the longitudinal magnetization change on the
right is much less than that on the left at the boundary.
The charge neutrality condition
$\int_{\frac{d}{2}}^\infty \delta\! n^L d z + \int^{-\frac{d}{2}}_{-\infty}
\delta\! n^L d z =0$
yields
\begin{equation}
\delta n^R_0=
-(l_1\delta n_{10}^L+l_2\delta n_{20}^L)/\lambda ,
\end{equation}
These two equations express the quantities on the right in terms of
quantities on the left. We now determine the amplitudes of these physical 
quantities by matching the boundary condition as in eq. (7). 


{\it Boundary conditions:}
The longitudinal 
magnetization current in the ferromagnet arriving at the interface
$I_M^L$is
equal to the magnetization current $J_M^t$
across the interfacce due to tunnelling because the term on the right hand side of eq. (7) is along the transverse direction.
For the longitudinal component, $I_m=J_m.$ 
The longitudinal magnitization tunnelling current
is equal to the difference of the
spin up and the spin down tunnelling current.
From standard calculations
of the tunnelling current\cite{Mahan} we get
$$J_M^t=\sum_s s|T_{ss'}|^2(\delta n_{Ls}(E+\delta \mu_s^L)
-\delta n_{Rs}(E+\delta \mu_s^R)).$$
Here $\delta\mu$ contains contributions from the electric potential due to the 
charges at the interface and that from the accumulation due to the bottleneck
effect. The change of the electron density of spin s can be related
to the change of the total charge and magnetization densities by
(we use units so that $\mu_B=1$): $\delta n_{s}=0.5(\delta n +s\delta M_x).$
The longitudinal magnetization density is the sum of contributions
from the solutions of the homogeneous
and the inhomogeneous equations:
$$\delta M_x=\delta m_x^i +\delta M_{x0}.$$
From eq. (14)
$J_M^L =\frac{(1- \beta^2) D_M^{\prime }}{l_{2}} \delta\! M^L_{20}.$
The inhomgeneous term $\delta m^i$
is uniform and does not contribute to the magnetization current $J_M^L$
\textbf{inside} the ferromagnet.
From $J_M^t=J_M^L$, we get
$$(1- \beta^2) D^{\prime L}_M\delta M_{20}/l_2=\sum_s s|T_{ss}|^2(\delta
n_{Ls}-n_{Rs}).$$
All variables of this equation can be written in terms of the two
independent variables $\delta M_{10,20}.$ 
%
Now $\delta n_{0Ls}=0.5[\delta n_{10,L}+\delta n_{20,L} +s(\delta
M_{10,L}+\delta M_{20,L}).]$
Using eq. (11) and (16) we get
\begin{equation}
(1-\beta ^{2})D_{M}^{\prime L}\delta
M_{20}/l_{2}=\sum_{s}s|T_{ss}|^{2}[\delta \!M_{10}^{L}(\xi
^{L}-1+s+l_{1}(\xi^{L}-1)/\lambda )+s\delta m_x^i].
\end{equation}%
This equation implies that
$\delta M_{20}$ is of the order of $c_t\delta M_{10}^L/c_m$
where $c_t$  ($c_m$) is the tunnelling (metallic) conductance.  $c_t$
much smaller than the metal conductance $c_m$. 
Thus $\delta M_{20}$ is much smaller than $\delta M_{10}.$

For an open circuit, the total charge tunnelling current is
zero. We get $J=\sum_{s}|T_{ss}|^{2}(\delta n_{Ls}-n_{Rs})=0.$
%
%
Substituting in the expresssions for the charge densities
and using the condition that $\delta M_{20}<<\delta M_{10}$, we get
\[
\sum_s |T_{ss}|^2[\delta\!M_{10}^L (\xi^L-1 +s+l_1(\xi^L-1) /\lambda)
+s\delta m_x^i ] =0.
\]

Solving this equation, we finally obtain
\begin{equation}
\delta \!M_{10}^{L}=-f \delta m^{i}_x
\end{equation}%
where $f=(\sum_{s}|T_{ss}|^{2}s)
/[\sum_{s}|T_{ss}|^{2}(\xi^{L}-1+s+l_{1}(\xi^{L}-1)/\lambda )].$
The corresponding charge is, from eq. (10), $\delta
\!n_{10}^{L}=(\xi^L-1)f\delta m^i_x.$
The charge and the magnetization densities
are proportional only to the {\bf ratio} of the conductances.
Hence they are not necessarily small for tunnel junctions.
As we emphasized before\cite{chui1}, this comes about because $\lambda<<l_{sf}.$

%

{\it Emf:}

The DC voltage is estimated as
the change of the mean chemical potential across the interface, given by
$\Delta V=0.5\sum_s \Delta(\delta\mu_s)=
0.5\sum_s \Delta(\delta n_s/N_s)$ where $N_s$ is the density of states.
This drop includes a contribution from an electric potential as well as a 
contributions from electron density changes due to bottleneck 
and electron-electron interaction effects.
This drop can be written as ($\delta M^L>>\delta M^R$)

$$\Delta V=0.25 (e\delta m^i_x/\mu_B)f
[(\xi^L-1)(1/N_+^{L}+2l_1/(\lambda^R N^R)+1/N_-^L)+1/N_+^L-1/N_-^L].$$

The longitudinal 
magnetization density is $\delta m_x^{i}=0.5(\delta m_{\perp}^{i})^{2}/(M_{0}v
)=0.5\theta^2M_0/v $ where $v$ is the atomic volume,
$\theta =\delta m^{i}/M_{0}$ is the precession angle. 
Hence
\begin{equation}
\Delta V=0.125 e\theta^2 M_0/(v\mu_B) f
[(\xi^L-1)(1/N_+^{L}+2l_1/(\lambda N^R)+1/N_-^L)+1/N_+^L-1/N_-^L].
\end{equation}

As expected, this d.c. voltage is proportional to $\theta^2,$
as is observed experimentally.
Most importantly, it is proportional only to a {\bf ratio} of the conductances.
Hence its magnitude is not small.
The factor f, as given after eq. (18), depends on the asymmetry between the 
majority spin and the minority spin conductances in the insulator. 
The larger the difference, the larger the value of $|f|$.
We next estimate the order of magnitude of $Delta V$.

We expect $M_0/(v\mu_B)$ to be of the order unity,
$e/N$ ( $N$ is the average density of states ) to be of the order of 0.1 volt.
Dependening on the asymmetry between the majority and the minority spin
tunnel conductances in the insulator, 
the value of f can range between 1 and 0.1.
Similarly, depending on the asymmetry between the majority and the 
minority spin conductances in the ferromagnet $\xi^L-1$ can range in 
value between 1 and 0.1;
$1/N_+-1/N_-$
to be of the order of $1/N$ to $0.1/N.$ 
Hence $\Delta V\approx (10^{-2}-10^{-4})\theta^2 volt.$
For $\theta\approx 0.1$, $\Delta V\approx (10^{-4}-10^{-6}) volt,$ 
in agreement with
the experimental results, which is larger than microvolts.
We next address the issue of damping.


{\it Damping:}
The loss can come from three sources: (1) from the interface inhomogeneity,
(2) from loss of the transverse magnetization current through the barrier,
(3) from loss of the longitudinal magnetization current. As we explain below,
these contributions have different dependence on the
external RF magnetic field. The contributions for the first two sources
to the damping coefficient are independent of the field strength;
that from the last source is proportional to the input power.
The contributions from the last two sources are inversely
proportional to the junction resistance and thus are much smaller
for tunnel junctions. 

We first estimate the loss connected with the longitudinal magnetization.
This loss is equal to $ \sum_{s}j_{s}^{2}r_{s}$ where $j_s$, $r_s$ is  
the current and junction resistance for spin s.
This is of the order of $(\delta m_x^i)^2 |T|^2$. Since $\delta m_x$
is proportional to the input power, this loss is proportional
to the power squared. Its contribution to the damping coefficient is
obtained by normalizing the loss by the energy density and hence is 
proportional to the power. Because this loss is proportional to $|T|^2$
its contribution is small 
for tunnel junctions. Similarly, we expect the transverse magnetization
current to incur a loss of the order of $(\delta m_{\perp}^i)^2|T|^2.$
Since $\delta m_{\perp}$ is proportional to the field,
this loss is propotional to the
power. Its contribution to the damping coefficient, again obtained by 
normalizing with espect to the energy density, is thus independent
of the power. This loss is also proportional to $|T|^2$
and will be small for tunnel junctions.

We next estimate the loss connected with the interface inhomogenity.
This requires knowledge of $\delta M_{30,40}$ which we now determine.
Again, we expect the transverse magnetization to be a sum of a term
that is the solution of the inhomogeneous equation ($\delta m^i$)
and terms that are solutions of the homogeneous equation ($\delta M_{3,4}$).
We calculate $\delta M_{3,4}$ using the boundary condition given by eq. (7).
From eq. (15) the transverse magnetization current at the boundary is
\[
{\hat{\mathbf{J}}_{\mathrm{M}}}^{L}=-\frac{D_{M}^{\prime }}{l_{3}}\mathbf{e}%
_{+}\delta \!M_{30}^{L}-\frac{D_{M}^{\prime }}{l_{4}}\mathbf{e}_{-}\delta
\!M_{40}^{L}
\]%
The pseudo spin current in eq. (7) is thus given by
$${\hat{\mathbf{I}}_{\mathrm{M}}}^{L}=-(D_{M}^{\prime }
+i\gamma JM_0)\delta M_{30}^L \mathbf{e}_{+}/l_3-(D_{M}^{\prime } 
-i\gamma JM_0)\mathbf{e}_{-} \delta \!M_{40}^{L}/l_4
$$



Eq. (7) also involves the tunnelling transverse current $J_m^t$. To evaluate
this we follow standard practice\cite{Mahan} 
and calculate the rate of change of the
transverse magnetization due to tunnelling.
%
%
We found that the tunnelling current for the transverse magnetization can be
written as
$J_{M+}^t=M_+^L(g_1+ig_2)+M_+^R(g_3+ig_4)$
where\cite{details}  $g_{1,2,3,4}$ are proportional to $|T|^2.$
A similar equation for $J_{M-}$ can be written down.
This shows that the contribution from the tunnelling current is
smaller than the other terms in eq.(7) and thus will be treated
by perturbation theory.
We finally obtain to lowest order $-I_{\bf M}^L=\gamma K_s{\bf M}_0
\times (\delta
{\bf m}^i+\delta {\bf M}_{0s}). $ Substituing in the
expression for $I_m$, this equation becomes
$$[\pm i D_{M}^{\prime } /(l_{\pm}\gamma K_s M_0 )-1- J/(l_{\pm}K_s)]\delta
M_{\pm 0}^L = \delta m^i_{\pm}.
$$
Here $l_+=l_3,$ $l_-=l_4$, $\delta M_+=\delta M_{30},$ $\delta M_-=
\delta M_{40}.$ 
As we go away from the interface,
the transverse magnetization density dies off exponentially. The total 
magnetization is given by
$\delta M_{\pm}l_{\pm}=\delta m^i_{\pm}/X_{\pm}$
where $X_{\pm}=
[\pm i D_{M}^{\prime } /(\gamma K_s M_0 )-l_{\pm}- J/K_s].$
The correction term due to the tunnelling magnetization current
is equal to $-X^{-1}I_m^t/(\gamma K_sM_0).$

The magnetic susceptibility, given by $(\delta M_{\pm}^L l/d_F+\delta m_{\pm})
/H_{1,\pm},$  becomes $\chi=\chi^0(1+X^{-1}/d_F).$
The additional damping comes from the imaginary part of  $\chi$ which now
contains a term proportional to $Re(\chi^0)Im X^{-1}/d_F.$ 
This term is proportional to the metallic "resistance" $D_M'$ which in turn 
comes from the spatially varying part of the magnetization induced by
the surface, as we have anticipated.
This contribution is not a function of the junction resistance
and will be of the same order of magnitude for multilayers as well as for 
tunnel barriers. 

In conclusion, we discussed in this paper
the voltage and the damping of an RF field in
ferromagnetic tunnel junctions. The voltage is controlled
by changes of the longitudinal magnetization whereas the
damping seems mainly associated with the transverse magnetization.
Additional sources that can induce transverse magnetization localized
near the interface can come from localized changes of the Hamiltonian
such as the surface anisotropy. 
The calculation in this paper can be trivially extended to junctions
with ferromagnets on both sides.
For junctions involving two ferromagnets on opposite sides
(F1-I-F2 or F1-F2), the interface anisotropy $K_s$ will contain a 
term from the dipolar interaction between F1 and F2. The loss will then
be a function of the orientation of the magnetizations of F1 and F2,
consistent with experimental results.

STC is supported in part by the DOE. We thank John Xiao for helpful
conversation.
\end{document}